\documentstyle[12pt]{article}
\begin {document}

\begin{center}
{\bf THE ROLE OF HADRONIZATION IN CHARM AND BEAUTY PRODUCTION}

\vspace{0.5cm}

Yu.M.Shabelski \\
Petersburg Nuclear Physics Institute, \\
Gatchina, St.Petersburg 188300 Russia \\

\end{center}

\vspace{0.5cm}

\begin{abstract}
We discuss the relative role of fragmentation and recombination
processes for heavy flavour hadron production in different
kinematical regions in high energy hadron-hadron and photon-hadron 
collisions. We predict several qualitative features which should be 
observed if our picture of heavy flavour production is consistent.

\end{abstract}

\vspace{2cm}

E-mail SHABELSK@THD.PNPI.SPB.RU

\newpage

\section{Introduction}

The investigation of heavy flavour production in high energy hadron
collisions is an important method for studying the quark-gluon
structure of hadrons and the mechanism of hadroproduction at high
energies.

The most popular and technically simplest approach is the so-called QCD
collinear approximation, or parton model (PM). The cross sections of QCD
subprocess are calculated usually in the leading order (LO), as well
as in the next to leading order (NLO) \cite{1,2,NDE,Beer,Beer1}. The
possibility to incorporate the incident parton transverse momenta is
referred to as $k_T$-factorization approach
\cite{CCH,CE,MW,CH,CC,HKSST}, or the theory of semihard interactions
\cite{GLR,LR,8,lrs,3,SS,BS}. In our previous papers \cite{our, our1} we
have presented a comparison of results obtained with the help of
$k_T$-factorization and the parton model.

However in both approaches we calculate the cross sections,
one-dimential distributions and correlations of the produced heavy
quarks, whereas experimentally we know these quantities for heavy
flavour hadrons. Of course, the total production cross sections for
heavy flavour quarks and hadrons are the same due to flavour
conservation, however the shape of distributions and correlations can
be different.

The aim of this paper is the discussion of these differences in 
different kinematical domains. We will give several qualitative
predictions which seem to be trivial. However their experimental check
is very important because their violation means that our to-day picture 
of hadronization is inconsistent. The preference of heavy flavours comes
from the fact that at not asymptotically high energy only one heavy
quark pair can be produced. So we have the explicite correspondence
between the calculated heavy quark and experimentally measured heavy
flavour hadron production.

\section{Fragmentation and recombination approach for secondary
production}

There exist two classes of the phenomenological models for hadronization
which account for two different processes: fragmentation of the produced 
quark into secondary hadron and recombination of the produced quark with 
some another quark into secondary hadron. 

Experimentally we know that both these processes exist. Quark 
fragmentation takes place in the case of heavy flavour production in 
$e^+e^-$ annihilation. On the other hand, only recombination processes 
of the produced heavy quark with valence quarks of incident hadrons
can explain (see, for example, \cite{Likh,TNKN,jdd}) the
experimental asymmetry \cite{E769,WA82,E769a,E791} in yields of leading
(favoured) and non-leading (unfavoured) $D$-mesons\footnote{The 
intrinsic charm idea \cite{VBH} is slightly different.} which is 
defined as
\begin{equation}
A(x) = \frac{d\sigma /dx (Leading) - d\sigma /dx (Non-leading)}
{d\sigma /dx (Leading) + d\sigma /dx (Non-leading)} \;,
\end{equation}
where "leading" hadrons have the common light quark with the incident
particle, and "non-leading" ones have no such quark.

The similar asymmetry was measured for charmed baryon $\Lambda^+_c$ to 
$\Lambda^-_c$ yields in $\pi^-$ nucleus interactions \cite{Ait}.

The total momentum distribution $D_H(p)$ of the produced heavy 
flavour hadron, at fixed value of transverse momenta is determined 
by the sum of the fragmentation and recombination processes,
$D_H^F(p)$ and $D_H^R(p)$,
\begin{equation}
D_H(p) = D_H^F(p) + D_H^R(p) \;.
\end{equation}

The principle difference of fragmentation and recombination approaches
is the difference of the ratio of the momentum of the produced heavy 
quark to secondary hadron. In the case of fragmentation the momentum 
of the hadron is smaller than the momentum of the quark, the momentum
distribution of heavy flavour hadron $D_H(p)$ is
\begin{equation}
D_H^F(p) = \int d^3p_1 D_Q(p_1) G(p/p_1) \;,
\end{equation}
where $D_Q(p_1)$ is the momentum distribution of heavy quark and $G(p/p_1)$
the fragmentation function of the quark into registrated hadron.

In the case of recombination of heavy quark $Q$ with light antiquark (or
diquark) $q$ we have the opposite situation.
\begin{equation}
D_H^R(p) = \int d^3p_1 d^3p_2 D_Q(p_1) D_q(p_2) \delta (p - p_1 - p_2) 
\;,
\end{equation}
and the momentum of secondary hadron is larger than the momentum of 
every quark. 

One can see that the fragmentation contribution into some secondary 
hadron distribution depends only on the momentum distribution of heavy 
quarks, whereas the recombination distribution depends both on the 
momentum distributions of heavy quarks and light antiquarks (or 
diquark). So the relative contribution of fragmentation and 
recombination processes in Eq.~(2) depends on the density of light 
antiquarks (diquarks) in the considered kinematical region.

Sometimes it was sayd that heavy flavour hadron can not be 
produced via recombination of a heavy quark with light antiquark/diquark 
because of significant difference in their average transverse momenta. 
However the average transverse momenta of secondaries (mesons) produced 
centrally at high energy are of the order of their masses, 
$\langle p_T \rangle \sim m_H$. On the other hand in a fast heavy flavour
meson most probably both, heavy and light quarks have almost equal
velocities\footnote{This configuration gives the main contribution, say,
in nonrelativistic quark model.}. Taking into account that the
constituent mass of heavy quark is parametrically larger that the
constituent mass of light quark, $m_Q \gg m_q$, we will find in heavy
flavour meson $H$ with standard transverse momentum $p_T \sim m_H$ a heavy
quark which carries $p_T \sim m_Q$, and a light quark which carries its
usual transverse momentum $p_T \sim m_q$.

As both fragmentation and recombination processes exist, the question is
--  which process is the most important for the heavy flavour hadron
production in the high energy hadron collisions, and how this can 
depends on the kinematical region. The experimental fact 
\cite{E769,WA92} is that the calculated Feynman-$x$ distribution of 
produced heavy quarks in $\pi p$ collisions at fixed target energies are 
in good agreement with the experimental distributions of produced 
$D$-mesons, see Fig. 5 in \cite{FMNR}. So we can say that in the 
averaged events with charm production the fragmentation and recombination 
processes in charm quark hadronization balanced each other.

However, this balance should be violated if we will consider Feynman-$x$
distribution of heavy flavoured mesons with some restriction in their
transverse momenta. For example, in events with heavy quark pair
production at comparatively small $p_T \sim m_Q$ the multiplicity of
light quarks is several times (in dependence on the initial energy)
larger than the multiplicity of heavy quarks. So here we have many
objects (antiquarks or diquarks) which can recombinate with heavy quark.
However, in the region of $p_T \gg m_Q$ the $p_T$-distributions of
light and heavy quarks should be practically the same, because both are
determined by the same QCD diagrams with scale, equal to large $p_T$
value. In this kinematical region the probability of recombination
should decrease (see examles in \cite{NS}), due to decrease of the 
relative density of light antiquarks/diquarks with the needed 
comparatively large transverse momentum, $D_q(p)$ in Eq.~(4), whereas 
the probability of fragmentation should be the same as at small $p_T$. 
So we can expect that the balance between recombination and 
fragmentation will be changed and the spectra of secondary heavy flavour 
hadrons should be more soft than the spectra of produced heavy quarks. 
The experimental check of this behaviour seems to be very interesting. 
By the way, possibly the dependence of the difference in 
$x_F$-distribution of heavy quarks and hadrons on the $p_T$ values can 
explain why $p_T$ distributions of heavy quarks are changed after $k_T$ 
kick and fragmentation (where $p_T$ values are comparatively large), 
whereas $x_F$-distributions (which are controlled by low $p_T$ values) 
becomes practically the same as before fragmentation \cite{Shab}.

At the same time, if the contribution of recombination for heavy flavour
hadrons with large $p_T$ decrease, we predict the decrease of the
asymmetry Eq.~(1) in the production of leading to non-leading
secondaries. Besides this, the measurement of the dependence of
asymmetry on the transverse momenta of secondaries gives information
about wave function of heavy flavour hadrons.

Very interesting behaviour is expected for the energy behaviour of the
asymmetry (1) in $ep$ collisions. As is well-known, at not very high 
initial energy the direct $\gamma p$ interactions dominate, when the
incident photon goes into $Q\bar{Q}$ pair with production another 
secondaries via soft parton shower. So the heavy quark pair and the 
proton remnant are in different hemispheres, and the difference of 
their rapidities increase with initial energy fast enough. In this 
configuration the probability of heavy quark recombination with proton
remnant is small and the asymmetry should decrease with the energy. 
However, with the growth of energy the resolved process starts to 
contribute more and more. Here heavy quark pair is produced in 
parton-parton collision via the hadron component of a photon. The cross 
sections of the last processes are determined by both parton structure 
functions of proton and photon. The photon structure functions are known 
not good enough, so the ratio of resolved to direct contributions can be 
changed several times, by using different sets of parton distributions 
in the photon, one can see several numerical examples in \cite{Shab1}.      

In the resolved photon process heavy quarks are produced in the
central region, similarly to the $pp$ case, so the asymmetry should be
practically the same. The direct interactions should give only small 
correction. So the measurement of the ratio of heavy flavour hadron
asymmetry in $\gamma p$ (in the proton fragmentation hemisphere)
and in $pp$ collisions at the same energy can estimate the ratio of 
resolved to direct processes. It seems to be also interesting to 
consider these ratios in different $p_T$ regions.

\section{Conclusion}

We have discussed the qualitative features of heavy quark hadronization.
It seems that in very general approach the hadronization mechanism
should depend on the transverse momentum of heavy quark/hadron. This
results in some qualitative predictions for the data. Probably the most 
important feature should be the decrease of the asymmetry, Eq.~(1) with 
the transverase momenta of heavy flavour hadrons. 

   The comparison of heavy flavour hadron asymmetries in $\gamma p$ and
$pp$ collisions at similar energies allows one to estimate the 
part of interactions with resolved photon.

Of course, all quantitative estimations here are model dependent 
\cite{Likh,NS,BJZ,Likh1}.

I am grateful to M.G.Ryskin for useful discussions.

%\newpage

\end{document}